\documentclass[a4paper,10pt,aps,pre,twocolumn,superscriptaddress,showpacs,amsmath,amssymb,nofootinbib]{revtex4}
\usepackage{graphicx,color}

\newcommand{\dd}{\mathrm{d}}
\newcommand{\td}[2]{\frac{\dd #1}{\dd #2}}
\newcommand{\pd}[2]{\frac{\partial #1}{\partial #2}}
\newcommand{\fd}[2]{\frac{\delta #1}{\delta #2}}
\newcommand{\mean}[1]{\langle #1 \rangle}
\newcommand{\Int}[1]{\int\dd #1\;}
\newcommand{\IInt}[3]{\int_{#2}^{#3}\dd #1\;}

\renewcommand{\vec}[1]{\mathbf #1}

\newcommand{\al}{\alpha}

\newcommand{\Gam}{\Gamma}

\newcommand{\eps}{\varepsilon}
\newcommand{\kap}{\kappa}
\newcommand{\lam}{\lambda}
\newcommand{\vhi}{\varphi}
\newcommand{\sig}{\sigma}

\newcommand{\ie}{\emph{i.e.}}
\newcommand{\eg}{\emph{e.g.}}

\newcommand{\im}{\text{i}}
\newcommand{\ra}{\rightarrow}
\newcommand{\x}{\vec r}
\newcommand{\Dr}{D_\text{r}}
\newcommand{\tu}{\tau_\text{r}}
\newcommand{\De}{D_\text{e}}
\newcommand{\vc}{v_\text{c}}
\newcommand{\xc}{\zeta_\text{c}}
\newcommand{\ac}{\al_\text{c}}
\newcommand{\bc}{\beta_\text{c}}

\newcommand{\drho}{\delta\rho}

\newcommand{\ad}{_\text{ad}}
\newcommand{\xs}{^\text{(sp)}}
\newcommand{\xb}{^\text{(bi)}}
\newcommand{\nois}{\boldsymbol\xi}
\newcommand{\kT}{k_\text{B}T}

\begin{document}

\title{Dynamical mean-field theory and weakly non-linear analysis for the
  phase separation of active Brownian particles}

\author{Thomas Speck}
\affiliation{Institut f\"ur Physik, Johannes Gutenberg-Universit\"at Mainz,
  Staudingerweg 7-9, 55128 Mainz, Germany}
\author{Andreas M. Menzel}
\author{Julian Bialk\'e}
\author{Hartmut L\"owen}
\affiliation{Institut f\"ur Theoretische Physik II,
  Heinrich-Heine-Universit\"at, D-40225 D\"usseldorf, Germany}

\begin{abstract}
  Recently, we have derived an effective Cahn-Hilliard equation for the phase
  separation dynamics of active Brownian particles by performing a weakly
  non-linear analysis of the effective hydrodynamic equations for density and
  polarization [Phys. Rev. Lett. \textbf{112}, 218304 (2014)]. Here we develop
  and explore this strategy in more detail and show explicitly how to get to
  such a large-scale, mean-field description starting from the microscopic
  dynamics. The effective free energy emerging from this approach has the form
  of a conventional Ginzburg-Landau function. On the coarsest scale, our
  results thus agree with the mapping of active phase separation onto that of
  passive fluids with attractive interactions through a \emph{global}
  effective free energy (mobility-induced phase transition). Particular
  attention is paid to the square-gradient term necessary for the dynamics. We
  finally discuss results from numerical simulations corroborating the
  analytical results.
\end{abstract}


\maketitle


\section{Introduction}

The separation of a suspension of passive colloidal particles into liquid and
vapor is a complex and rather well-studied phenomenon~\cite{ande02}. For phase
separation to occur, sufficiently strong attractions between particles have to
be present. This is well understood from the perspective of thermodynamics:
The potential energy gained by the suspension forming the dense phase
compensates for the loss of entropy.

Recently, a separation into dense and dilute regions has also been reported
for ``active'' suspensions composed of self-propelled colloidal
particles~\cite{theu12,pala13,butt13}. For a brief perspective on this
phenomenon see Ref.~\citenum{bial14}, and
Refs.~\citenum{roma12,marc13,elge14,menzel2015tuned} for more general recent
reviews on various aspects of active matter. By constantly converting external
energy into directed motion, such systems of self-propelled particles can be
driven into a non-equilibrium steady state. Steady dynamic states of
orientationally ordered collective motion can
arise~\cite{vicsek1995novel,tone95,evans2011orientational,
  menzel2012soft,ferrante2013elasticity,menzel2013traveling,alarcon2013spontaneous,
  ferrante2013collective,weber2014defect,menzel2014active}, but also swirling
and turbulent-like situations were
observed~\cite{saintillan2008pattern,wensink2012meso,wensink2012emergent,gros14}. There
are different possibilities how to provide the energy input in experiments:
light, if sufficiently strong, can create a temperature gradient leading to
self-thermophoresis~\cite{jian10,breg14}, or lead to the local demixing of a
binary solvent~\cite{butt12}. In most experiments, however, energy is provided
chemically, \eg, through the decomposition of hydrogen peroxide~\cite{paxt06}
or the release of stored ions~\cite{rein13}. All of these mechanisms are based
on a symmetry breaking on the particle level. In the case of spherical Janus
particles~\cite{walther2008janus} it is provided by different surface
properties of typically two distinct hemispheres.

What is intriguing from a fundamental perspective is that the phase separation
of active colloids strongly resembles the phase separation in a passive
suspension, but occurs also for \emph{purely repulsive} particles. This has
been demonstrated convincingly in computer simulations of a minimal model by a
number of groups~\cite{yaou12,redn13,bial13,fily14}. This minimal model of
active Brownian particles incorporates the two basic physical ingredients:
volume exclusion and persistence of motion, \ie, particles interact via
repulsive potentials (or hard-core exclusion) and every particle has an
orientation along which it ``swims'' at constant speed. The particle
orientations evolve independently and stochastically. The microscopic reason
for particle accumulation is simple (see, \eg, the kinetic model in
Ref.~\citenum{redn13}): Due to the persistence of the self-propelled motion
and the excluded volume, particles block each other on the time scale that it
takes for orientations to decorrelate. If the mean collision time is shorter
(the suspension is sufficiently dense) clustering ensues. Simulations with
idealized boundary conditions show that hydrodynamic interactions due to the
solvent can modify these time scales~\cite{zott14,mata14}. On the other hand,
experiments~\cite{butt13} indicate that for colloidal swimmers the basic
scenario is robust, and we will neglect hydrodynamic interactions in the
following.

Even if the microscopic mechanism is known, the collective large-scale and
phase behavior is still highly non-trivial. Tailleur and Cates have been the
first to realize that the phenomenon of phase separation in active systems can
be explained by the dependence of the particle mobility on local
density~\cite{tail08,cate13,cate14}. The resulting theory is referred to as
``mobility induced phase separation'' (MIPS). Within this framework it has
been demonstrated that the temporal evolution of the coarse-grained density
can be mapped onto that of an effective bulk free energy. Good agreement
between particle-resolved simulations of active Brownian particles and the
numerical solution of the coarse-grained density has been
demonstrated~\cite{sten13,sten14}. In analogy to the mean-field treatment of
phase separation in passive suspensions (with attractive forces), binodal and
spinodal lines are identified from the minima and inflection points of the
effective bulk free energy, respectively, from which a schematic phase diagram
is constructed. Both lines merge at a single critical point. The resulting
free energy involves only the density, a result that is based on the
elimination of the polarization through neglecting temporal and spatial
derivatives (a detailed discussion follows in Sec.~\ref{sec:ad}). In
particular, this treatment produces a bulk free energy that does not contain a
term that penalizes sharp interfaces. It is thus not able to describe the
dynamics of domain formation and coarsening. In order to cure this
shortcoming, Cates and coworkers have argued that the dynamics has to be
augmented by gradient terms that are not derivable from a free
energy~\cite{sten13,witt14}.

The purpose of the present paper is to follow the complete path from the
microscopic dynamics to the large-scale Cahn-Hilliard equation~\cite{cahn58}
and to give a comprehensive derivation of the results obtained in
Ref.~\citenum{spec14}. To this end we derive effective hydrodynamic equations
and perform a weakly non-linear analysis~\cite{cros93}. Our results refine the
MIPS scenario in two ways: (i)~Our systematic derivation leads to a
square-gradient term that completes the bulk free energy to the conventional
Ginzburg-Landau form. (ii)~We clarify the validity of a description in terms
of the density alone and show that it can strictly hold only close to the loss
of linear stability. The physical reason is that this instability occurs on
length scales that are larger than the persistence length of the directed
motion and thus allow us to study the coarse-grained density alone. The
microscopic ``trapping'' of particles due to their directed motion then enters
as an effective attraction. Quenching deeper into the two-phase region, the
non-equilibrium nature of active suspensions will become evident again and the
coupled evolution of density and orientation has to be considered.

Our paper is organized as follows: In Sec.~\ref{sec:mf}, we discuss active
Brownian particles as a minimal model for self-propelled disks. Starting from
the full many-body dynamics, we sketch the derivation of the effective
hydrodynamic equations of motion following Ref.~\citenum{bial13}. In
Sec.~\ref{sec:ad}, we explore the consequences of the \emph{adiabatic
  approximation} of the hydrodynamic equations leading to an effective
equilibrium theory for the density alone. We then describe the weakly
non-linear analysis in Sec.~\ref{sec:weak}, the ramifications of which are
discussed in Sec.~\ref{sec:discussion}. Conclusions and outlook are given in
Sec.~\ref{sec:conc}.


\section{Mean-field theory}
\label{sec:mf}


\subsection{Minimal model for self-propelled disks}

The model we study consists of $N$ identical, interacting disks with free
diffusion coefficient $D_0$ moving in a periodic box with area $A$. Each disk
has an orientation $\vec e_k=(\cos\vhi_k,\sin\vhi_k)^T$ that undergoes free
rotational diffusion with diffusion coefficient $\Dr$. We consider both
diffusion coefficients to be constant and eliminate them through rescaling
time $t\mapsto t/\Dr$ and length $\x\mapsto\ell\x$, where
$\ell\equiv\sqrt{D_0/\Dr}$. Neglecting hydrodynamic interactions, the coupled
equations of motion become
\begin{equation}
  \label{eq:lang}
  \dot\x_k = -\nabla U + v_0\vec e_k + \nois_k,
\end{equation}
where $U=\sum_{k<k'}u(|\x_k-\x_{k'}|)$ is the potential energy (in units of
the thermal energy) with pair potential $u(r)$, and the noise correlations due
to the solvent read
\begin{equation}
  \mean{\nois_k(t)\nois^T_{k'}(t')} = 2\mathbf{1}\delta_{kk'}\delta(t-t').
\end{equation}
The orientational angles obey
\begin{equation}
  \mean{\dot\vhi_k(t)\dot\vhi_{k'}(t')} = 2\delta_{kk'}\delta(t-t').
\end{equation}
We are thus left with two dimensionless parameters defining a non-equilibrium
state point: the number density $\bar\rho\equiv N\ell^2/A$ and the free speed
$v_0$.


\subsection{Derivation}

An equivalent representation of the equations of motion Eq.~\eqref{eq:lang} is
given by the Smoluchowski equation
\begin{equation}
  \label{eq:sm}
  \partial_t\psi_N = \sum_{k=1}^N\left\{ \nabla_k\cdot\left[(\nabla_k U) -
      v_0\vec e_k + \nabla_k\right] + \pd{^2}{\vhi_k^2} \right\} \psi_N
\end{equation}
for the joint probability $\psi_N(\{\x_k,\vhi_k\};t)$ of particle positions
and their orientations. This representation is convenient since it allows for
systematic approximations. Following Ref.~\cite{bial13}, we aim to derive a
closed equation of motion for the one-point particle density
\begin{equation}
  \label{eq:psi:1}
  \psi_1(\x_1,\vhi_1;t) \equiv \Int{\x_2\cdots\dd\x_N} \Int{\vhi_2\cdots\dd\vhi_N}
  N\psi_N
\end{equation}
integrating out all other particles. The \emph{local density}
\begin{equation}
  \label{eq:rho:def}
  \rho(\x,t) \equiv \Int{\vhi} \psi_1(\x,\vhi;t)
\end{equation}
corresponds to the probability of finding a particle at position $\x$ at time
$t$.

\begin{figure}[t]
  \centering
  \includegraphics{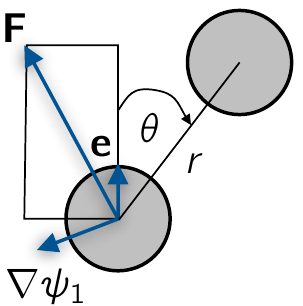}
  \caption{Sketch of the force decomposition. The total average force $\vec F$
    is approximately decomposed into a component along the particle
    orientation $\vec e$ and along the local one-point density gradient
    $\nabla\psi_1$. Also shown is one of the surrounding particles at distance
    $r$ enclosing the angle $\theta$ with the particle orientation.}
  \label{fig:sketch}
\end{figure}

Inserting Eq.~\eqref{eq:sm} into the time derivative of Eq.~\eqref{eq:psi:1},
we obtain
\begin{equation}
  \label{eq:mf:pre}
  \partial_t\psi_1 = -\nabla\cdot[\vec F + v_0\vec e\psi_1-\nabla\psi_1]
  + \partial^2_\vhi \psi_1
\end{equation}
dropping the particle index. The force $\vec F(\x,\vhi;t)$ is due to
interactions of a (fixed) particle with its surrounding particles averaged
over their accessible positions and orientations. It thus couples to the
two-point density
\begin{multline}
  \psi_2(\x_1,\vhi_1,\x_2,\vhi_2;t) \equiv \Int{\x_3\cdots\dd\x_N} \\
  \times\Int{\vhi_3\cdots\dd\vhi_N} N(N-1)\psi_N.
\end{multline}
Even in a homogeneous suspension of active particles, the force
$\vec F(\x,\vhi;t)$ does not vanish. Rather, there is a force imbalance on
each particle due to the directional motion, which implies a higher
probability to find another particle in front of it (looking along its
orientation) than behind. One finds
$F\equiv\vec e\cdot\vec F=-\rho\zeta\psi_1$ with the force imbalance
quantified through the coefficient~\cite{bial13}
\begin{equation}
  \label{eq:zeta}
  \zeta \equiv \IInt{r}{0}{\infty} r[-u'(r)] \IInt{\theta}{0}{2\pi}
  \cos\theta g(r,\theta)
\end{equation}
with pair distribution function $g(r,\theta)$. Here, $\theta$ is the angle
between the orientation of a particle and the displacement vector to another
particle at a distance $r$, see Fig.~\ref{fig:sketch}. For an inhomogeneous
density $\rho(\x)$ also $\zeta$ depends in principle on the position
$\x$. Close to the dynamical instability of the homogeneous density profile we
can neglect this spatial dependence and in the following we assume
$\zeta=\zeta(\bar\rho,v_0)$ to be a state function.

We decompose the force
\begin{equation}
  \label{eq:F}
  \vec F = (\vec e\cdot\vec F)\vec e + \delta\vec F \approx
  (\vec e\cdot\vec F)\vec e + F_\parallel\nabla\psi_1
\end{equation}
into a component along the particle orientation due to the force imbalance and
a component $\delta\vec F$ perpendicular. The later describes an ``evasive''
motion leading to an effective diffusion. Hence, as a closure we only keep the
projection of $\delta\vec F$ onto the density gradient $\nabla\psi_1$. This is
a good approximation as long as $|\delta\vec F|\sim 1$ is much smaller
compared to $\bar\rho\zeta\sim v_0\gg1$ [cf. Eq.~\eqref{eq:xc}]. Rearranging
$\nabla\psi_1\cdot\vec F$ leads to the formal expression\footnote{There is a
  typo in Ref.~\cite{bial13} below Eq.~(11) (it should read $|\nabla\psi_1|^2$
  in the expression for $D$). Consequently, Eq.~\eqref{eq:F} is not an
  expansion in the density gradient but follows from the different magnitudes,
  $\vec e\cdot\vec F\gg F_\parallel$.}
\begin{equation}
  F_\parallel \approx \frac{[\nabla\psi_1-(\vec e\cdot\nabla\psi_1)\vec
    e]\cdot\vec F}{|\nabla\psi_1|^2}.
\end{equation}
Inserting Eq.~\eqref{eq:F} into Eq.~\eqref{eq:mf:pre}, the mean-field
evolution equation for the joint probability of position and orientation thus
reads
\begin{equation}
  \label{eq:mf:joint}
  \partial_t\psi_1 = -\nabla\cdot[v(\rho)\vec e - \De\nabla]\psi_1 +
  \partial_\vhi^2\psi_1
\end{equation}
with effective diffusion coefficient $\De\equiv 1-F_\parallel>0$ and effective
speed
\begin{equation}
  \label{eq:v}
  v(\rho) \equiv v_0 - \rho\zeta
\end{equation}
depending on the local density.

For a homogeneous suspension, $\rho(\x,t)=\bar\rho$ is a constant. Assuming
$\De$ also to be constant, Eq.~\eqref{eq:mf:joint} formally corresponds to the
stochastic evolution of a single self-propelled particle. We can then
calculate the mean-square displacement and from that the long-time diffusion
coefficient~\cite{hows07}
\begin{equation}
  \label{eq:D}
  D(\bar\rho,v_0) = \De(\bar\rho) + \frac{1}{2}[v(\bar\rho)]^2.
\end{equation}
This relation has indeed been confirmed in several numerical studies of active
suspensions~\cite{bial13,sten13}, and therefore in the remainder we will treat
$\De(\bar\rho)$ as spatially uniform and independent of the speed. This
constitutes our final approximation closing Eq.~\eqref{eq:mf:joint}. Clearly,
$\De$ corresponds to the diffusion coefficient of the passive suspension
($v_0=0$).

To briefly conclude, Eq.~\eqref{eq:mf:joint} describes the evolution of the
active suspension on a coarse-grained level, into which the effects of
microscopic particle interactions enter through two effective parameters:
$\De$ and the force imbalance $\zeta$. Within the theory, every state point
$(\bar\rho,v_0)$ is fully characterized by these two parameters. However, the
theory cannot make predictions about their values, for which we would have to
make further assumptions or measure them in particle-resolved computer
simulations (as has been done in Ref.~\citenum{bial13}).

\subsection{Hydrodynamic equations}
\label{sec:hydro}

For the evolution of the local density Eq.~\eqref{eq:rho:def} one finds
\begin{equation}
  \label{eq:rho}
  \partial_t\rho = -\nabla\cdot[v(\rho)\vec p - \De\nabla\rho]  
\end{equation}
using Eq.~\eqref{eq:mf:joint}. This equation expresses number conservation
with a particle current that is given by a diffusive term $-\De\nabla\rho$ and
a current $v\vec p$ proportional to the \emph{polarization} or orientational
order parameter
\begin{equation}
  \vec p(\x,t) \equiv \Int{\vhi} \vec e \psi_1(\x,\vhi;t).
\end{equation}
For $\vec p\neq0$, particles in a coarse-grained volume have a preferred
orientation leading to a net particle current. This orientation evolves
according to
\begin{equation}
  \label{eq:p}
  \partial_t\vec p = -\nabla P(\rho) + \De\nabla^2\vec p - \vec p
\end{equation}
with pressure
\begin{equation}
  \label{eq:press}
  P(\rho) \equiv \frac{1}{2}v(\rho)\rho
\end{equation}
resulting from the directed motion of the particles. Inserting
Eq.~\eqref{eq:v}, for sufficiently large $\zeta$ this pressure becomes a
non-monotonic function of density. Hence, for local densities
$\rho>v_0/(2\zeta)$ the density gradient, $\nabla P(\rho)=P'(\rho)\nabla\rho$,
changes sign so that particles migrate towards denser regions. The second term
in Eq.~\eqref{eq:p} is akin to a ``viscosity term'' and the last term
describes the local relaxation due to the rotational diffusion.

While we have derived Eqs.~\eqref{eq:rho} and~\eqref{eq:p} starting from the
full microscopic Smoluchowski equation~\eqref{eq:sm}, the formal structure of
the result is not surprising since it has to reflect macroscopic conservation
laws and symmetries. Note that run-and-tumble dynamics leads to the same
hydrodynamic equations (cf. Ref.~\citenum{cate13} and see
Appendix~\ref{sec:rnt}). Although colloidal self-propelled particles of course
move in a solvent, the decision to neglect hydrodynamic interactions places
the resulting effective hydrodynamic theory into the field of what is
sometimes called ``dry active matter''~\cite{marc13}. Similar equations are,
\eg, obtained in the Toner-Tu continuum treatment~\cite{tone95} of polar
active systems~\cite{bert06,thur13}. In that case the alignment of
orientations leads to nonlinear terms of the polarization $\vec p$ in
Eq.~\eqref{eq:p} and thus to dynamical collective behavior.

\subsection{Linear stability}

\begin{figure}[t]
  \centering
  \includegraphics{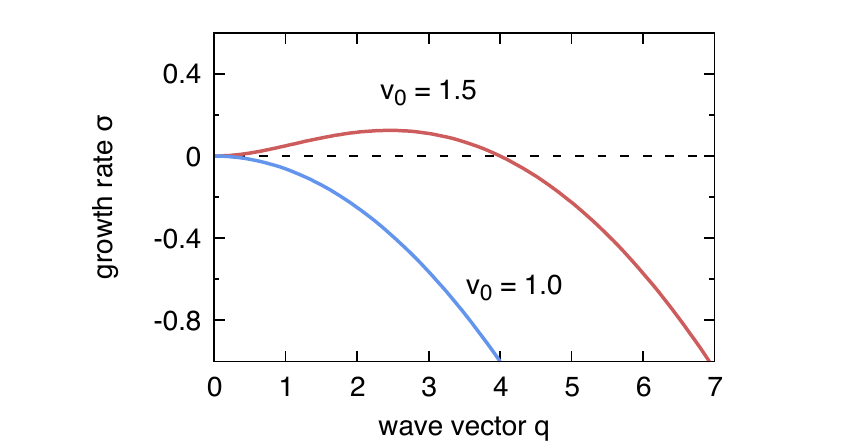}
  \caption{Dispersion relation $\sig(q)$ for two speeds $v_0$ below and above
    the critical speed $\vc\simeq1.15$ corresponding to $\bar\rho\zeta=1$ and
    $v_\ast=1$. For $v_0>\vc$ large-scale perturbations (\ie, small wave
    vectors $q>0$) become unstable.}
  \label{fig:disp}
\end{figure}

Clearly, any constant density with $\vec p=0$ is a stationary solution of
Eqs.~\eqref{eq:rho} and~\eqref{eq:p}. To study the stability of the
homogeneous density $\bar\rho$ with respect to small perturbations, we set
$\rho=\bar\rho+\drho$ and rewrite the equations of motion~\eqref{eq:rho}
and~\eqref{eq:p}
\begin{gather}
  \label{eq:rho:d}
  \partial_t\drho = -\al\nabla\cdot\vec p + \De\nabla^2\drho +
  \zeta\nabla\cdot(\vec p\drho), \\
  \label{eq:p:d}
  \partial_t\vec p = -\beta\nabla\drho + \De\nabla^2\vec p - \vec p +
  \zeta\drho\nabla\drho.
\end{gather}
Here, we have separated the non-linear terms and defined the coefficients
\begin{equation}
  \label{eq:coeff}
  \al \equiv v(\bar\rho) = v_0 - \bar\rho\zeta, \quad 
  \beta \equiv \frac{1}{2}(v_0 - 2\bar\rho\zeta).
\end{equation}
Dropping the non-linear terms and inserting solutions of the form $\drho,\vec
p\propto e^{\sig(q)t+\im\vec q\cdot\x}$, we obtain the dispersion relation
\begin{equation}
  \label{eq:disp}
  \begin{split}
    \sig(q) &= -\frac{1}{2} - \De q^2 + \frac{1}{2}\sqrt{1-4\al\beta q^2} \\
    &= -(\De+\al\beta)q^2 - (\al\beta)^2q^4 + \mathcal O(q^6),
  \end{split}
\end{equation}
which quantifies the growth rate of a perturbation with wave vector $q$ (see
Fig.~\ref{fig:disp}). Hence, for $\De+\al\beta<0$ a smooth perturbation of the
homogeneous density on small $q$ does not decay anymore but grows, leading to
a dynamical instability. Solving this condition implies an \emph{instability
  line} $\vc(\bar\rho)$ of critical speeds (for simplicity we only consider
the smaller solution) such that for $v_0>\vc$ the suspension becomes linearly
unstable. The growth rate $\sig(q)$ is maximized for the wave length
\begin{equation}
  \label{eq:q:max}
  q_0^2 = \frac{1}{4}\left(\frac{1}{\al\beta} - \frac{\al\beta}{\De^2} \right),
\end{equation}
which will thus dominate the morphology at early stages after the onset of the
instability.


\section{Adiabatic approximation}
\label{sec:ad}

For slowly varying fields, we neglect the temporal derivative as well as the
viscosity term in Eq.~\eqref{eq:p} to obtain the adiabatic solution
\begin{equation}
  \label{eq:p:adia}
  \vec p \approx \vec p\ad = -\nabla P = -\frac{1}{2}\nabla(v\rho).
\end{equation}
Inserting Eq.~\eqref{eq:v} for the density-dependent effective speed $v(\rho)$
there are two possibilities to rewrite this expression, which we now discuss.

\subsection{Relaxation of density}

First, we push the gradient to the right,
\begin{equation}
  \vec p\ad = -\frac{1}{2}(v_0-2\rho\zeta)\nabla\rho.
\end{equation}
We have assumed that $\zeta$ is constant, which holds for the homogeneous
density profile. Inserting $\vec p\ad$ into Eq.~\eqref{eq:rho} leads to the
simple evolution equation for the density
\begin{equation}
  \label{eq:rho:adia}
  \partial_t\rho = \nabla\cdot[\mathcal D(\rho)\nabla\rho]
\end{equation}
with collective diffusion coefficient
\begin{equation}
  \label{eq:D:coll}
  \mathcal D(\rho) = \De + \frac{1}{2}(v_0-\rho\zeta)(v_0-2\rho\zeta).
\end{equation}
The global density at which the diffusion coefficient switches its sign is
given by the condition $\mathcal D(\bar\rho)=0$. It signals the onset of a
dynamical instability at which the homogeneous density profile
$\rho(\x,t)=\bar\rho$ becomes (linearly) unstable and small perturbations
start to grow. Using the coefficients introduced in Eq.~\eqref{eq:coeff} we
find $\mathcal D(\bar\rho)=\De+\al\beta$ and therefore $\sig(q)=-\mathcal
D(\bar\rho)q^2$ demonstrating that it is the same instability discussed in the
previous section.

\subsection{Free energy}
\label{sec:free}

Alternatively, to illustrate the concept of an effective free energy as
advocated by Cates and Tailleur~\cite{cate13,tail08}, we cast the evolution
equation for the density into a form that involves the functional derivative
of a potential function, which thus can be interpreted as an effective free
energy. This is achieved by pulling out the Nabla operator,
\begin{equation}
  -v\vec p\ad = \frac{1}{2}\nabla\left[ v_0^2\rho -
    \frac{3}{2}v_0\zeta\rho^2 + \frac{2}{3}\zeta^2\rho^3 \right].
\end{equation}
Inserting this result into Eq.~\eqref{eq:rho}, we now find the evolution
equation
\begin{equation}
  \label{eq:rho:f}
  \partial_t\rho = -\nabla\cdot(v\vec p\ad-\De\nabla\rho) =
  \nabla^2\fd{F\ad}{\rho}
\end{equation}
implying the functional
\begin{equation}
  \label{eq:ff:ad}
  F\ad[\rho] = \Int{^2\x} f\ad(\rho(\x))
\end{equation}
with bulk free energy density
\begin{equation}
  f\ad(\rho) = \frac{1}{2}\left(\De+\frac{v_0^2}{2}\right)\rho^2 -
  \frac{1}{4}v_0\zeta\rho^3 + \frac{1}{12}\zeta^2\rho^4.
\end{equation}
We will refer to this function as the adiabatic free energy density. It has
the typical form of a Landau function often encountered in the study of
critical phenomena and phase transitions~\cite{chaikin}.

In order to discuss the phase diagram following from the adiabatic solution,
we rewrite the bulk free energy density as a symmetric function plus a linear
term,
\begin{multline}
  \label{eq:f:ad}
  f\ad(\rho) = f\ad(\rho_0) + \mu(\rho-\rho_0) \\
  + \frac{1}{2}\left(\De - \frac{v_0^2}{16}\right)(\rho-\rho_0)^2 +
  \frac{1}{12}\zeta^2(\rho-\rho_0)^4,
\end{multline}
with
\begin{equation}
  \rho_0 \equiv \frac{3v_0}{4\zeta}, \qquad
  \mu \equiv \rho_0\left(\De + \frac{v_0^2}{8} \right).
\end{equation}
The bulk free energy density becomes a non-convex function for speeds
\begin{equation}
  v_0\geqslant 4\sqrt\De\equiv v_\ast,
\end{equation}
where $v_\ast(\bar\rho)$ depends on the global density. As the speed is
increased beyond $v_\ast$, the suspension enters the two-phase region. A
common-tangent construction to minimize the bulk free energy density in the
non-convex region leads to
\begin{equation}
  \label{eq:f:min}
  \left.\pd{f}{\rho}\right|_{\rho\xb_\pm} = \mu
\end{equation}
with the coexisting densities
\begin{equation}
  \label{eq:bi}
  \rho\xb_\pm = \rho_0 \pm \frac{\sqrt 3}{4\zeta}\sqrt{v_0^2-v_\ast^2}.
\end{equation}
Consequently, the coefficient $\mu$ represents the chemical potential, which
according to Eq.~\eqref{eq:f:min} is equal for the two coexisting phases of
densities $\rho\xb_\pm$ identifying the \emph{binodal}. Since
$f\ad''(\rho)=\mathcal D(\rho)$, the inflection points defining the mean-field
\emph{spinodal}
\begin{equation}
  \label{eq:sp}
  \rho\xs_\pm = \rho_0 \pm \frac{1}{4\zeta}\sqrt{v_0^2-v_\ast^2}
\end{equation}
coincide with the limit of linear stability as expected.

\begin{figure}[t]
  \centering
  \includegraphics{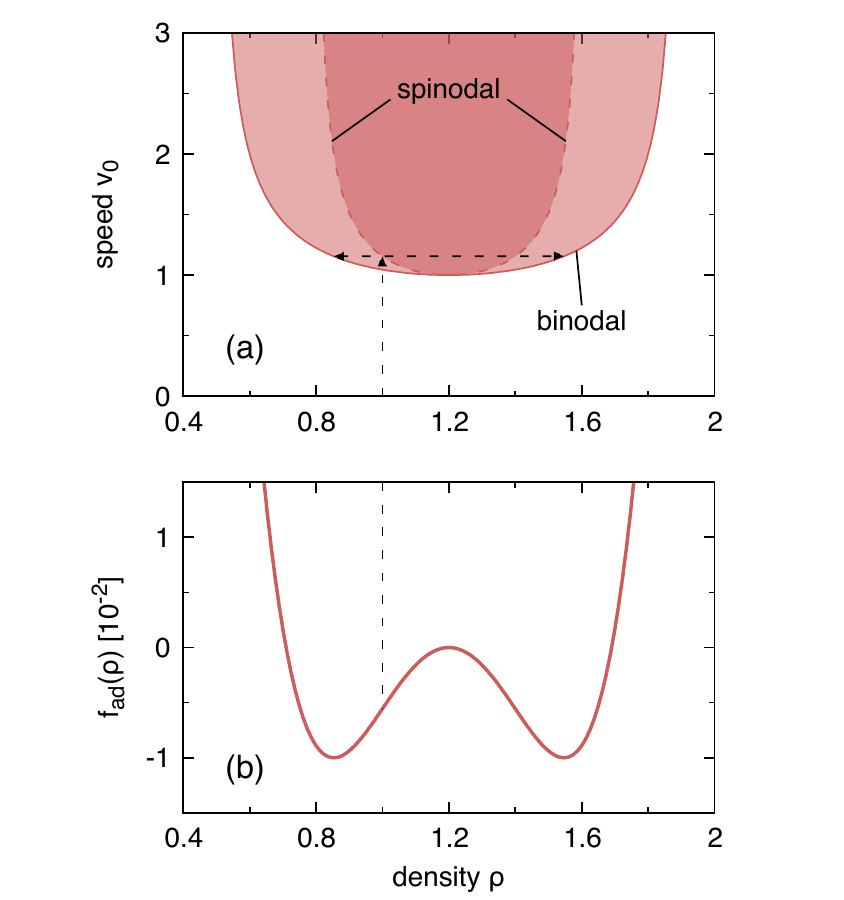}
  \caption{Schematic phase diagram for the adiabatic mean-field free energy
    density Eq.~\eqref{eq:f:ad} using $\zeta(v_0)=(3v_0)/(4\bar\rho_\ast)$
    with critical point at $\bar\rho_\ast=1.2$ and $v_\ast=1$ (corresponding
    to constant $\De=1/16$). (a)~Spinodal (dashed) and binodal (solid) lines
    in the $(\rho,v_0)$-plane. Indicated is the behavior for global density
    $\bar\rho=1$: Increasing the propulsion speed $v_0$ (vertical arrow), the
    homogeneous profile loses linear stability when reaching the spinodal at
    $\vc$. For any quenched $v_0$, the coexisting densities are predicted by
    the points on the binodal (horizontal arrows for $\vc$). (b)~The
    corresponding tilted free energy density $f\ad(\rho)-\mu(\rho-\rho_0)$ for
    $v_0=\vc$. The continuation of the dashed line from the upper panel
    indicates the inflection point for $\bar\rho=1$.}
  \label{fig:ad}
\end{figure}

The physical picture following from this discussion is thus that of an active
suspension undergoing a phase separation into a dense and a dilute phase, see
Fig.~\ref{fig:ad} for an illustration. Although we study a system that is
incessantly driven away from thermal equilibrium, its large-scale evolution is
apparently that of an effective equilibrium suspension, where the speed $v_0$
plays the role of an inverse temperature. In particular, the existence of a
free energy-like functional guarantees a unique stationary state in which this
functional becomes minimal,
\begin{equation}
  \td{F}{t} = \Int{^2\x} \fd{F}{\rho}\pd{\rho}{t} = -\Int{^2\x}
  \left|\nabla\fd{F}{\rho}\right|^2 \leqslant 0
\end{equation}
inserting Eq.~\eqref{eq:rho:f}. In qualitative agreement with numerical
simulations~\cite{yaou12,redn13,sten13}, the mean-field theory predicts a
binodal enclosing the two-phase region and a spinodal within this region. The
condition
\begin{equation}
  \rho_0 = \frac{3v_\ast}{4\zeta_\ast} = \bar\rho_\ast
\end{equation}
with $\zeta_\ast\equiv\zeta(\bar\rho_\ast,v_\ast)$ defines the critical point
$(\bar\rho_\ast,v_\ast)$ at which binodal and spinodal meet.

To conclude this section we comment on two points: First, the functional
Eq.~\eqref{eq:ff:ad} does not contain a gradient term penalizing sharp
interfaces between low and high density phase, and we will come back to this
point in Sec.~\ref{sec:close}. Second, the pressure $P(\rho)$
[Eq.~\eqref{eq:press}] does not have to be equal in the coexisting phases. The
interface will consist of particles pointing into the dense phase (otherwise
particles will leave the interface back into the dilute phase). Hence, the
orientation $\vec p$ does not vanish in the interface allowing for a jump of
$P(\rho)$ following Eq.~\eqref{eq:p:adia}. In addition to the pressure
$P(\rho)$, from a free energy density $f(\rho)$ one can derive the
``thermodynamic'' pressure
\begin{equation}
  \label{eq:press:th}
  P_f(\rho) = -\pd{(af)}{a} = -f(\rho) + \rho f'(\rho)
\end{equation}
with $P_f\neq P$, where $a$ is a (small) area with locally homogeneous density
$\rho\propto1/a$. This pressure now becomes equal for the coexisting densities
Eq.~\eqref{eq:bi}.


\section{Weakly non-linear analysis}
\label{sec:weak}

\subsection{Dominating mode}

Employing the adiabatic solution Eq.~\eqref{eq:p:adia} allows to reduce the
original hydrodynamic equations to a single equation of motion for the density
alone, which, moreover, can be cast into a form involving an effective free
energy. We now attempt to study the behavior of the hydrodynamic equations in
a more systematic way employing a small expansion parameter~\cite{spec14}. To
this end we will consider the state points $(\bar\rho,\vc)$ along the
instability line with force imbalance coefficients
\begin{equation}
  \label{eq:xc}
  \bar\rho\xc = \frac{3}{4}\vc - \frac{1}{4}\sqrt{\vc^2-v_\ast^2}
\end{equation}
following from the condition $\De+\ac\bc=0$. We study speeds
\begin{equation}
  v_0 = \vc(1+\eps)
\end{equation}
close to the instability line with $|\eps|\ll1$. We expand the coefficients
$\al(v_0)=\ac+\al_1\eps+\mathcal O(\eps^2)$ and
$\beta(v_0)=\bc+\beta_1\eps+\mathcal O(\eps^2)$ into Taylor series assuming
that they are analytic functions of the speed $v_0$. To leading order
$\al\beta\approx-\De+\sig_1\eps$ with new coefficient
\begin{equation}
  \label{eq:sig}
  \sig_1 \equiv \ac\beta_1 + \al_1\bc.
\end{equation}
Hence, the fastest growing wave vector Eq.~\eqref{eq:q:max} behaves as
$q_0\sim\sqrt\eps$ and the growth rate of structures with this wave vector
becomes $\sig(q_0)\approx-\sig_1q_0^2\eps\sim\eps^2$. In the following it will
be more convenient to employ non-negative $\eps\geqslant0$ and let
$\sig_1\mapsto\pm|\sig_1|$ so that the sign of $\sig_1$ determines on which
side of the instability line we are: for $\sig_1>0$ (small) fluctuations decay
while for $\sig_1<0$ the suspension has become unstable.

We now aim to derive an equation of motion for the density fluctuations on the
scale of the dominating mode. Since these fluctuations evolve on the length
$1/q_0$ and grow with time scale $1/\sig(q_0)$ we rescale length and time
leading to
\begin{equation}
  \label{eq:rescale}
  \partial_t\mapsto\eps^2\partial_t, \qquad
  \nabla\mapsto\sqrt\eps\nabla,
\end{equation}
which we plug into Eqs.~\eqref{eq:rho:d} and~\eqref{eq:p:d}.

\subsection{Close to the critical point}
\label{sec:close}

In a first step, we expand the local density and orientation as
\begin{gather}
  \label{eq:c:sq}
  \rho = \bar\rho + \sqrt\eps c + \eps c^{(1)} + \eps^{3/2}c^{(3/2)} + \cdots, \\
  \vec p = \eps\vec p^{(1)} + \eps^{3/2}\vec p^{(3/2)} + \cdots.
\end{gather}
To lowest order the magnitude of density fluctuations is thus $\sim\sqrt\eps$,
\emph{viz.} the response is $\drho\propto\eps^{1/2}$ as expected close to a
critical point with mean-field exponent $\tfrac{1}{2}$. The expansion form for
$\vec p$ has been chosen to match powers. Plugging all expansions back into
the hydrodynamic equations together with Eq.~\eqref{eq:rescale}, we collect
terms of the same order $\eps$. To lowest order, we find
$\vec p^{(1)}=-\bc\nabla c$ and therefore
\begin{equation}
  \label{eq:c:fst}
  0 = (\De+\ac\bc)\nabla^2 c,
\end{equation}
which is fulfilled for any perturbation $c(\x,t)$ since the expression in
brackets corresponds to the instability condition and thus vanishes.

To next order we find $\vec p^{(3/2)}=-\bc\nabla c^{(1)}+\xc(c\nabla c)$
leading to
\begin{equation}
  \label{eq:c:sec}
  0 = -2g\nabla\cdot(c\nabla c) = -g\nabla^2 c^2
\end{equation}
with another coefficient
\begin{equation}
  \label{eq:g}
  g \equiv \frac{1}{2}\xc(\ac+\bc) = \frac{1}{4}\xc\sqrt{\vc^2-v_\ast^2}
  \geqslant 0.
\end{equation}
For non-vanishing $c$ this condition is fulfilled only for $g=0$, which
implies $\vc=v_\ast$. Hence, the expansion Eq.~\eqref{eq:c:sq} for the density
fluctuations is only valid in an $\eps$-environment of the critical point
$(\bar\rho_\ast,v_\ast)$.

\begin{figure*}[t]
  \centering
  \includegraphics{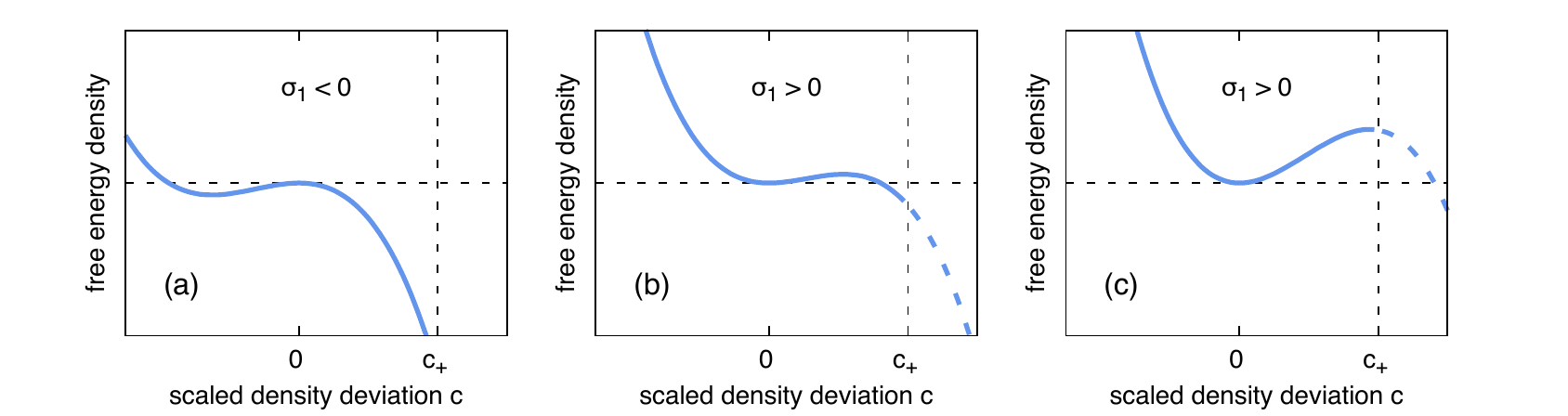}
  \caption{Possible shapes of the free energy density Eq.~\eqref{eq:f}: (a)~In
    the linearly unstable region for $\sig_1<0$. Due to volume exclusion there
    is a maximal $c_+$ (vertical dashed line) at which the density has a
    minimum. (b,c)~For $\sig_1>0$ two cases are possible: (b)~the uniform
    density profile corresponding to $c=0$ is metastable (the second minimum)
    or (c)~the uniform density profile is globally stable.}
  \label{fig:free}
\end{figure*}

Assuming $g=0$, at the next order we finally obtain an equation of motion
\begin{equation}
  \partial_t c = \sig_1\nabla^2 c + \xc^2\nabla\cdot(c^2\nabla c) -
  \De^2\nabla^4 c = \nabla^2\fd{F_{1/2}}{c}
\end{equation}
for the density fluctuations. The right hand side can be expressed as the
functional derivative of a potential function (an effective free energy)
\begin{equation}
  \label{eq:ff}
  F_{1/2}[c] = \Int{^2\x} \left[ \frac{\De^2}{2}|\nabla c|^2 + f_{1/2}(c) \right]
\end{equation}
with bulk term
\begin{equation}
  \label{eq:f:sq}
  f_{1/2}(c) = \frac{1}{2}\sig_1 c^2 + \frac{1}{12}\xc^2 c^4.
\end{equation}
In contrast to the adiabatic solution Eq.~\eqref{eq:ff:ad}, the expansion in
$\eps$ now includes the customary square-gradient term.

As the final step, we restore the density through inserting
$c=(\rho-\bar\rho)/\sqrt\eps$ and reverting the scaling
Eq.~\eqref{eq:rescale}. While the gradient term is invariant, the bulk
contribution to the free energy density becomes
\begin{equation}
  \eps^2 f_{1/2} = \frac{1}{2}\sig_1\eps(\rho-\bar\rho)^2 
  + \frac{1}{12}\xc^2(\rho-\bar\rho)^4.
\end{equation}
In order to show that this expression for the bulk is equivalent to
Eq.~\eqref{eq:f:ad} obtained from the adiabatic solution, we first consider
\begin{equation}
  \De - \frac{v_0^2}{16} = \De - \frac{v_\ast^2}{16}(1+\eps)^2 \approx
  -\frac{v_\ast^2}{8}\eps
\end{equation}
up to linear order in $\eps$. On the other hand, we obtain
\begin{equation}
  \sig_1 = -\frac{v_\ast^2}{8}
\end{equation}
inserting the force imbalance coefficient Eq.~\eqref{eq:xc} at the critical
point $\vc=v_\ast$. This demonstrates that the adiabatic solution $f\ad$
coincides with the result $f_{1/2}$ for the bulk free energy density of a more
systematic expansion close to the critical point. In addition, the later route
also yields a square-gradient term that describes the cost of creating density
inhomogeneities. This term is important in the coarsening dynamics and allows
to predict an interfacial tension from the mean-field theory. Note that such a
square-gradient term can already be anticipated from the dispersion relation
Eq.~\eqref{eq:disp} from the $q^4$ term with $(\ac\bc)^2=\De^2$. The
coefficient defines a length $\De$ that can be interpreted as an effective
interaction range.

\subsection{Away from the critical point}

We found in the previous section that an expansion in density fluctuations of
order $\sim\sqrt\eps$ only holds for $\vc=v_\ast$, \ie, at the critical
point. To lift this restriction and to be able to move along the instability
line we need to satisfy Eq.~\eqref{eq:c:sec} while at the same time $g\neq 0$.
To achieve this within the $\eps$-expansion of the density one has to demand
that the leading contribution is of order $\sim\eps$ (instead of the larger
$\sim\sqrt\eps$). Effectively, this reduces the magnitude of density
fluctuations. We, therefore, arrive at the expansion employed in
Ref.~\cite{spec14},
\begin{gather}
  \label{eq:c:lin}
  \rho = \bar\rho + \eps c + \eps^2 c^{(2)} + \cdots, \\
  \vec p = \sqrt\eps[\eps\vec p^{(1)} + \eps^2\vec p^{(2)} + \cdots].
\end{gather}
As before, we collect terms of the same order. At lowest order we again find
Eq.~\eqref{eq:c:fst}. The next order already leads to the equation of motion
\begin{equation}
  \label{eq:c:eqm}
  \partial_t c = \sig_1\nabla^2 c - 2g\nabla\cdot(c\nabla c) -
  \De^2\nabla^4 c = \nabla^2\fd{F}{c}
\end{equation}
for the density fluctuations away from the homogeneous profile, where
\begin{equation}
  \label{eq:p:2}
  \vec p^{(2)} = -\bc\nabla c^{(2)} - \beta_1\nabla c + \De\nabla^2\vec p^{(1)}
  + \xc c\nabla c
\end{equation}
has been used. The form of the free energy functional coincides with
Eq.~\eqref{eq:ff}. In particular, we obtain the same square-gradient term as
before. However, the bulk free energy density now reads
\begin{equation}
  \label{eq:f}
  f(c) = \frac{1}{2}\sig_1 c^2 - \frac{1}{3}gc^3.
\end{equation}
Here the previously introduced coefficients $\sig_1$ and $g$ become the
coefficients of the square and cubic term, respectively.

The striking feature of Eq.~\eqref{eq:f} is that a $c^4$ term stabilizing the
dense phase is missing. Nevertheless, due to volume exclusion, a real
suspension will not collapse into a single point but reach the predicted
phase-separated state as demonstrated in experiments and simulations. This
mechanism of volume exclusion works on length scales corresponding to the
particle size implying $c<c_+$ with $c_+$ corresponding to the maximal
density. The damping thus arises from couplings to scales that are not
included in the systematic expansion at the above order. Thus the reason for
the missing $c^4$ term is the scale separation between the particle size and
the length over which density fluctuations are coarse-grained. The different
possible shapes of the free energy density $f(c)$ depending on the system
parameters are sketched in Fig.~\ref{fig:free}. Even without the $c^4$ term,
we can derive a differential equation that describes the instability line, see
Appendix~\ref{sec:diff}.


\section{Discussion}
\label{sec:discussion}

\subsection{Off-critical quenches}

The two local free energy densities $f_{1/2}(c)$ and $f(c)$ for the two
different scalings can be derived from the same global free energy. To
understand this and the role of the asymmetric $c^3$ term in the free energy
density Eq.~\eqref{eq:f} it is instructive to recall the situation (for
passive suspensions) with a symmetric free energy density
$f(\vhi)=\tfrac{a}{2}\vhi^2+\tfrac{\kap}{4}\vhi^4$ with order parameter
$\vhi\propto\rho-\bar\rho_\ast$. The Cahn-Hilliard equation models the
evolution after a quench from a stable homogeneous density $\vhi=-c_0$ into
the two-phase region, where $c_0=0$ implies a quench through the critical
point. For an off-critical quench, we set $\vhi=c-c_0$ with $c$ the deviations
away from the initial homogenous density as before. The Cahn-Hilliard equation
then reads
\begin{equation}
  \partial_t c = \nabla^2[(a+3\kap c_0^2)c-3\kap c_0c^2+\kap c^3] - \De^2\nabla^4c
\end{equation}
dropping constant terms which vanish due to the spatial derivative. Cleary,
this result becomes Eq.~\eqref{eq:c:eqm} with $g=3\kap c_0$ and
$\sig_1=a+g^2/(3\kap)$ when adding a repulsive term $\tfrac{1}{4}\kap c^4$ to
$f(c)$. The coefficient $a(v_0)$ should be a monotonically decreasing function
of speed $v_0$. It changes sign at $v_0=v_\ast$ and becomes negative for
$v_0>v_\ast$. A given global density $\bar\rho$ determines $\vc=\vc(\bar\rho)$
and $g=g(\vc)$ [Eq.~\eqref{eq:g}]. Exactly on the spinodal $\sig_1=0$ holds,
which implies $a=-g^2/(3\kap)$ for the quadratic coefficient of the global
free energy. Note that at $\vc=v_\ast$ we have $g=0$ and, therefore, we
recover the function $f_{1/2}(c)$ [Eq.~\eqref{eq:f:sq}] if in addition we set
$\kap(v_\ast)=\frac{1}{3}\zeta_\ast^2$. Hence, $f(c)$ extends the lower order
solution $f_{1/2}(c)$ to higher speeds $v_0$ but with undetermined coefficient
$\kap(v_0)$, which does not follow from the $\eps$-expansion.

\subsection{Nucleation behavior close to the spinodal}
\label{sec:nuc}

The Cahn-Hilliard equation has been derived originally to describe
\emph{spinodal decomposition}, the homogeneous, barrierless onset of phase
separation throughout the system in response to a quench beyond the spinodal
bounding the instable region. Interestingly, for the spinodal itself the
Cahn-Hilliard equation predicts a change from a continuous to a discontinuous
transition (see, \emph{e.g.} Ref.~\citenum{novi85}).

To discuss this effect in the present context of active Brownian particles, we
symmetrize and rescale the free energy density Eq.~(\ref{eq:f}) (including the
repulsive $c^4$ term) leading to the scaling functions
\begin{equation}
  \label{eq:f:scale}
  \begin{split}
    \tilde f_\pm(\eta) &\equiv \frac{\kap}{|\sig_1|^2}[f(c)-f(c_0)] \\
    &= \tilde\mu_\pm\eta + \frac{1}{2}(\pm 1-\Gam)\eta^2 + \frac{1}{4}\eta^4
  \end{split}
\end{equation}
with scaled density $\eta\equiv\sqrt{\kap/|\sig_1|}(c-c_0)$, where
$c_0=g/(3\kap)$ as in the previous section. These functions depend on the
single parameter
\begin{equation}
  \Gam \equiv \frac {gc_0}{|\sig_1|} = \frac {g^2}{3\kap|\sig_1|} \geqslant 0
\end{equation}
combining the three coefficients $\sig_1$, $g$, and $\kappa$, with irrelevant
$\tilde\mu_\pm(\Gam)$ akin to a chemical potential. Corresponding to the sign
of $\sig_1$, $\tilde f_-$ describes the effective free energy density in the
unstable and $\tilde f_+$ in the linearly stable region.

\begin{figure}[t]
  \centering
  \includegraphics{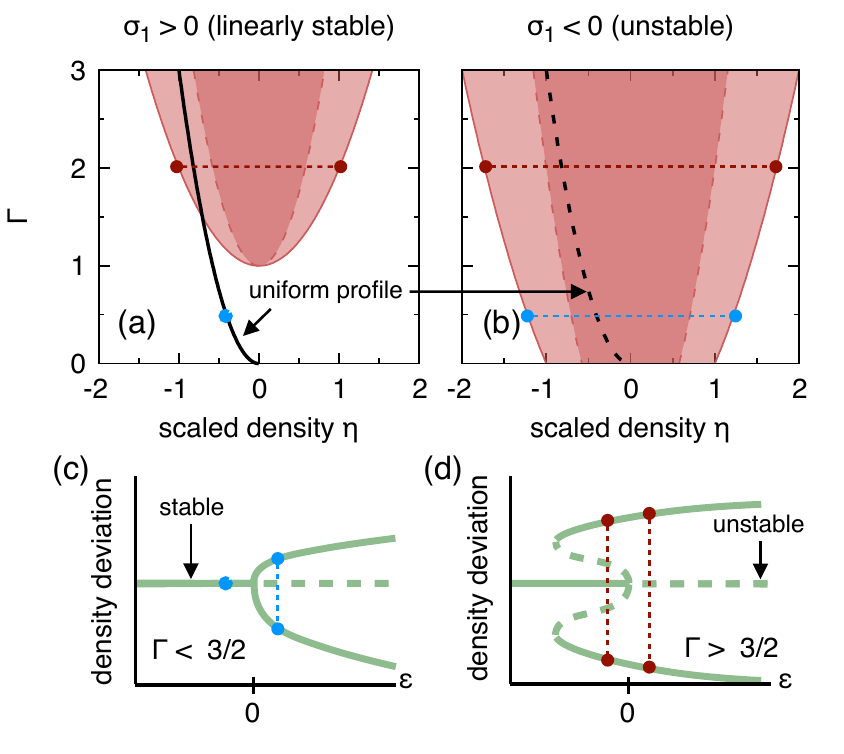}
  \caption{Schematic ``phase diagrams'' obtained from the scaling functions
    (a)~$\tilde f_+$ and (b)~$\tilde f_-$, see Eq.~\eqref{eq:f:scale}. Shown
    are coexisting densities (binodals, outer solid lines) and the limits of
    linear stability (spinodals, inner dashed lines). For $\sig_1>0$ the
    uniform density (black line) is linearly stable (it becomes metastable for
    $\Gam>\tfrac{3}{2}$) and for $\sig_1<0$ it is always
    unstable. (c,d)~Sketch of the bifurcation diagrams showing the deviation
    from the uniform density (the ``amplitude'', solid lines are linearly
    stable and dashed lines unstable) \emph{vs.} the control parameter $\eps$
    for fixed $\Gam$. The diagrams of (a) and (b) correspond to $\pm|\eps|$ on
    both sides of the bifurcation point as indicated by the connected dots for
    two representative values of $\Gam$: (c)~Supercritical pitchfork
    bifurcation for $\Gam<\tfrac{3}{2}$ corresponding to a continuous
    transition. (d)~Subcritical bifurcation for $\Gam>\tfrac{3}{2}$
    corresponding to a discontinuous transition.}
  \label{fig:phase}
\end{figure}

The parameter $\Gam$ is defined along the spinodal. In particular, the
critical point $(\bar\rho_\ast,v_\ast)$ corresponds to $\Gam=0$ and going away
from this point along the spinodal, the value of $\Gam$ increases. In
Fig.~\ref{fig:phase} the resulting ``phase diagram'' is shown in the
$(\eta,\Gam)$ plane. Due to the scaling employed in the previous section, the
functions $\tilde f_\pm(\eta)$ correspond to fixed (small) $\eps$ and we now
need to distinguish the two cases shown in Fig.~\ref{fig:phase}(a) and (b)
depending on which side of the spinodal the system resides. Shown are the
binodals calculated from the common tangent construction and the spinodals
corresponding to the inflection points, cf. Sec.~\ref{sec:free}. The uniform
density corresponds to setting $c=0$, \ie, $\eta=-\sqrt{\Gam/3}$. For
$\sig_1>0$ the uniform density is linearly stable but becomes metastable for
$\Gam>\tfrac{3}{2}$ (it crosses the binodal). This implies that for
$\Gam<\tfrac{3}{2}$ crossing the spinodal (jumping from $\sig_1>0$ to
$\sig_1<0$) corresponds to a continuous transition. In contrast, for
$\Gam>\tfrac{3}{2}$ the transition becomes discontinuous.

It is instructive to also consider the bifurcation diagrams shown in
Fig.~\ref{fig:phase}(c,d). Here we show schematically the solutions of the
amplitude equation, which correspond to the extrema of the effective free
energy. Plotted is the density with respect to the homogeneous density as a
function of $\eps=v_0/\vc-1$ for fixed $\Gam$. For any $|\eps|>0$ close to the
instability at $\eps=0$, the (scaled) solutions can be directly read off
Fig.~\ref{fig:phase}(a) for $\eps<0$ ($\sig_1>0$) and Fig.~\ref{fig:phase}(b)
for $\eps>0$ ($\sig_1<0$) as indicated for two values of $\Gam$. These
correspond to a continuous and discontinuous transition, respectively.

\subsection{Non-local speed}

Within the weakly non-linear analysis an integrable square-gradient term
stabilizing domains appears naturally. To study phase separation kinetics,
Cates and coworkers have followed a different route and posit that the active
Brownian particles sample the density on a length scale $\lam$ larger than the
interparticle spacing~\cite{sten13,witt14}. To lowest order the speed then
becomes a non-local function $v(\hat\rho)$ with
$\hat\rho=\rho+\lam^2\nabla^2\rho$ such that
\begin{equation}
  \label{eq:non-local}
  v(\hat\rho) \approx v(\rho) + v'(\rho)\lam^2\nabla^2\rho.
\end{equation}
Plugging such a non-local speed into the adiabatic solution
Eq.~\eqref{eq:p:adia} produces the desired square-gradient term but,
strikingly, would lead to an equation of motion for the density that also
involves non-integrable terms, \ie, it is not longer representable as the
functional derivative of an effective free energy. Apparent consequences are
discussed in Ref.~\citenum{witt14}.

Replacing the speed Eq.~\eqref{eq:v} with the expression
Eq.~\eqref{eq:non-local}, we can again study the systematic $\eps$-expansion
of the hydrodynamic equations~\eqref{eq:rho} and~\eqref{eq:p} close to the
instability line. While the lowest order of the $\eps$-expansion remains
unchanged, the non-local speed modifies the expression Eq.~\eqref{eq:p:2}
\begin{equation}
  \vec p^{(2)} \mapsto \vec p^{(2)} +
  \frac{1}{2}\bar\rho\xc\lam^2\nabla^2(\nabla c)
\end{equation}
for the orientation, where we have used $v'(\rho)=-\zeta$. Consequently, the
bulk free energy density Eq.~\eqref{eq:f} remains the same and the only effect
is that the coefficient of the square-gradient term is replaced by
\begin{equation}
  \De^2 \mapsto \De^2 + \frac{1}{2}\ac\bar\rho\xc\lam^2.
\end{equation}
Since the additional term is positive, this corresponds to a larger
interaction range as one would intuitively expect.

In summary, non-integrable terms were introduced when employing the adiabatic
approximation. They do not appear in a systematic treatment. Instead, the
systematic treatment naturally produces the well-known squared density
gradient term in the effective free energy functional. As a consequence, close
to the instability line and at onset (mean-field) phase separation kinetics of
active suspensions is predicted to not qualitatively differ from that of
passive suspensions \emph{even for a non-local speed}.

\subsection{Numerical simulations}
\label{sec:num}

\begin{figure*}[t]
  \centering
  \includegraphics{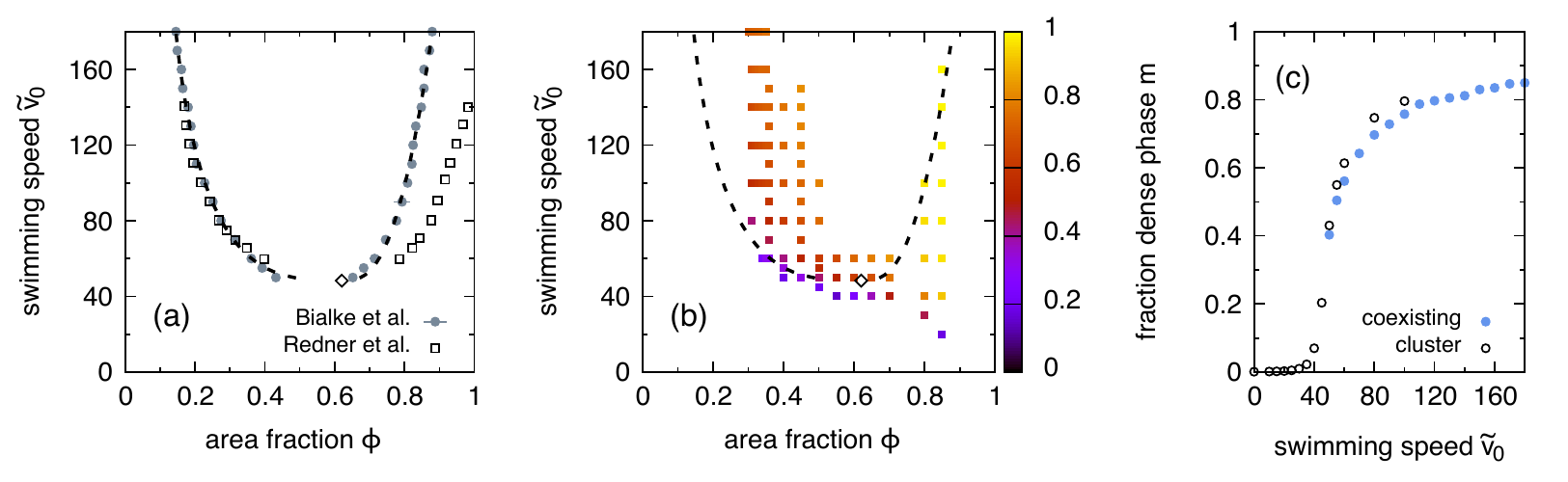}
  \caption{(a)~Coexisting densities for a model suspension, see text for
    details. Shown are ($\bullet$) the values from Bialk\'e \textit{et
      al.}~\cite{bial15} and ($\square$) Redner \textit{et
      al.}~\cite{redn13}. The dashed lines are fits of Eq.~\eqref{eq:gutt}
    with ($\diamond$) indicating the extrapolated position of the critical
    point. (b)~Fitted binodal and critical point as in (a) together with the
    average fraction $m$ of particles in the largest domain (scale bar on the
    right, from Ref.~\citenum{spec14}). Only state points with $m\geqslant0.1$
    are shown. (c)~Comparison of $m$ at $\phi=0.5$ for two methods: ($\circ$)
    from a cluster analysis to identify dense domains and ($\bullet$) from the
    coexisting densities via Eq.~\eqref{eq:m}.}
  \label{fig:coex}
\end{figure*}

Although there are already quite a few numerical studies of two-dimensional
active Brownian particles~\cite{yaou12,redn13,bial13,fily14,spec14,solo14a},
the determination of the full phase diagram is still an open task. As shown,
the mapping to a free energy strictly holds only close to the instability
line. Moreover, mean-field treatments are known to yield incorrect
quantitative predictions close to critical points due to a diverging
correlation length and the ensuing large fluctuations. These fluctuations make
numerical sampling notoriously difficult in the vicinity of a continuous
transition. This problem is already present in passive suspensions but appears
to be even more severe in active suspensions due to two reasons:
(i)~fluctuations are even larger and (ii)~advanced sampling techniques for
systems out of equilibrium are not (yet) available.

We examine the data from Refs.~\cite{redn13,spec14,bial15} obtained from
particle-resolved simulations of discoid self-propelled disks with diameter
$a$ interacting via the purely repulsive Weeks-Chandler-Andersen
potential~\cite{week72}
\begin{equation}
  u(r) =
  \begin{cases}
    4\epsilon[(\sig/r)^{12} - (\sig/r)^6] + \epsilon & (r < 2^{1/6}\sig) \\
    0 & (r \geqslant 2^{1/6}\sig)
  \end{cases}
\end{equation}
with potential strength $\epsilon$. We set $\sig=2^{-1/6}(a/\ell)$ so that
repulsive interactions are only present for overlapping particles. In the
following, we will present numerical results using the global area fraction
$\phi\equiv(a/\ell)^2\pi\bar\rho/4$, which is simply proportional to
densities. In principle, the rotational diffusion coefficient $\Dr$ is a free
parameter of the model. However, for self-propelled colloidal particles it is
reasonable to assume (see, \eg, supplementary materials of
Ref.~\citenum{butt13}) that the no-slip boundary condition still holds, which
implies $\Dr=3D_0/a^2$. This additional assumption relates the two lengths
$a=\sqrt{3}\ell$. For better comparison, in the following we report the
original numerical values for the dimensionless speed $\tilde v_0=\sqrt 3v_0$.

The accurate determination of coexisting densities in numerical simulations is
subtle. Redner~\textit{et al.} have extracted these densities from
histograms~\cite{redn13}. In Ref.~\citenum{bial15} a different strategy is
followed, in which a non-square simulation box with periodic boundaries is
employed. In such a geometry a ``slab'' of the dense phase forms, which aligns
with the shorter box edge. This allows a very accurate determination of
density profiles and the coexisting densities. The result for $N=4,900$ disks
at global area fraction $\phi=0.5$ using $\epsilon=100\kT$ is shown in
Fig.~\ref{fig:coex}(a). We have obtained data for $N=10,000$, which agree with
the coexisting densities for the smaller system demonstrating that there is no
finite-size dependence. Also shown is the data from Ref.~\citenum{redn13}, who
have used $\epsilon=\kT$ and $N=15,000$. The low-density branches agree very
well while for the high densities the interaction strength plays a more
important role. Clearly, the system with the higher $\epsilon$ is less
compressible. Still, both studies agree qualitatively. Note that below
$\tilde v_0\lesssim60$ fluctuations are so strong that a reliable
determination of coexisting densities is not possible anymore.

We have fitted the coexisting densities (the binodal) of Ref.~\citenum{bial15}
with the functional form
\begin{equation}
  \label{eq:gutt}
  \frac{\phi_\pm}{\phi_\ast} = 1 + a_\pm\left(\frac{v_0}{v_\ast}-1\right)
  + b_\pm \left(\frac{v_0}{v_\ast}-1\right)^{1/3}
\end{equation}
inspired by the famous Guggenheim plot~\cite{gugg45}. Here, $a_\pm$ and
$b_\pm$ are fit parameters, and $\phi_\pm$ are the coexisting area
fractions. This fit works surprisingly well over a wide range of speeds and
allows to extrapolate the position of the putative critical point:
$\phi_\ast\simeq0.62$ and $\tilde v_\ast\simeq48$. In Fig.~\ref{fig:coex}(b),
the fitted coexisting densities are overlaid by results from instantaneous
quenching: The passive suspension ($v_0=0$) is equilibrated at a given global
density. The suspension is then quenched to the final speed $v_0$ and relaxed
to the steady state. A cluster analysis is performed for the recorded
configurations to identify dense domains of $N_\text{c}$ particles, from which
the averaged fraction $m\equiv\mean{N_\text{c}}/N$ of particles in the largest
cluster (\ie, dense domain) is extracted. Note that only state points with
$m\geqslant0.1$ are shown, \ie, the largest domain contains at least 10\% of
all particles.

Fig.~\ref{fig:coex}(b) is compatible with the proposed mapping to passive
phase separation. Below area fraction $\phi\simeq0.3$ no spontaneous phase
separation is observed, which thus roughly indicates the location of the
spinodal. There is a larger region around the putative critical point where
spontaneous phase separation seems to occur even below the extrapolated
binodal, which, however, could be interpreted as a finite-size effect. In a
finite system, large fluctuations appear as dense domains but would not lead
to full phase separation in a larger system. However, this does not explain
the observed phase separation at higher area fraction $\phi\geqslant0.7$. At
these high densities details of the interparticle interactions and the
associated particle length scale may become important, which is not captured
by the coarse-grained point of view of the weakly nonlinear analysis.

Previously in Ref.~\cite{bial13} we have reported data that would suggest a
continuous transition (for $\phi=0.4$ and $\phi=0.5$). In line with this
observation, in Ref.~\cite{spec14} an abrupt onset of hysteresis for area
fraction $\phi<0.32$ has been observed, which thus agrees with the mean-field
prediction of a change from continuous to discontinuous (see
Sec.~\ref{sec:nuc}). Whether this observed change is a numerical artefact due
to the vicinity to a single critical point (as in passive phase separation) or
a genuine non-equilibrium effect remains to be investigated.

For completeness, we note that for a single dense domain (as observed at
intermediate system sizes) a simple relation between $m$ and the coexisting
densities exists via the lever rule,
\begin{equation}
  \label{eq:m}
  m(v_0) \approx \frac{1/\phi_- - 1/\phi}{1/\phi_- - 1/\phi_+}.
\end{equation}
This is demonstrated in Fig.~\ref{fig:coex}(c). The result for $m$ obtained
from the cluster analysis is slightly larger, which may be attributed to the
interfacial particles being counted towards the dense phase.


\section{Conclusions}
\label{sec:conc}

To summarize, we have followed a systematic route from the microscopic
dynamics to the large-scale Cahn-Hilliard equation to address the phase
behavior of active Brownian particles. In Ref.~\cite{bial13}, starting from
the microscopic dynamics, effective hydrodynamic equations have been derived
for the temporal evolution of (weakly perturbed) density and average
orientation. Two parameters enter these equations: the free speed $v_0$ and
the global density $\bar\rho$. All details of the particle interactions are
contained in the single function $\zeta=\zeta(\bar\rho,v_0)$, which quantifies
the force imbalance due to the interplay between repulsive forces and directed
motion.

The hydrodynamic equations exhibit a dynamical instability at
$\vc=\vc(\bar\rho)$, which can be determined through a linear stability
analysis~\cite{yaou12,bial13,fily14}. To go beyond the linear regime, we have
performed a weakly non-linear analysis~\cite{cros93} close to the instability
line using as the small expansion parameter $\eps=v_0/\vc-1$. This approach
yields an equation of motion for the density alone, which is formally
equivalent to a Cahn-Hilliard equation. It involves a local effective free
energy. We have discussed two expansions holding for different ranges of the
speed $\vc$, which together agree with the scenario of mapping active to
passive phase separation \emph{close to the loss of linear stability}. What
happens further away from the instability line? Including higher orders of the
expansion is of course possible but will lead to a large number of additional
terms. Already at the next order the non-equilibrium nature of the active
suspension will become manifest since a description in terms of density
fluctuations alone is not possible anymore and the time evolution of the
polarization has to be included. Hence, there seems to be no real advantage in
going to higher orders since one can also study the original hydrodynamic
equations.

One should stress that we have described a mean-field scenario for the
one-point density. Stationary two-point correlations are an input to the
theory and higher order correlations are neglected. Gaussian noise could be
added, which accounts for uncorrelated fluctuations but of course does not
restore the missing correlations. For passive suspensions it is well known
that mean-field free energies have to be regularized by the Maxwell
construction to be thermodynamically valid. Moreover, experimentally and in
simulations no sharp loss of linear stability is observed and the ``spinodal''
line is a pure mean-field concept~\cite{bind07}. Still, one might wonder why
for active Brownian particles a mean-field description nevertheless seems to
reproduce domain morphologies of particle-based simulations to such a good
degree~\cite{sten13}. It thus remains to be tested numerically to which extend
the theory outlined here is valid. Some evidence has been discussed in
Sec.~\ref{sec:num} but more detailed numerical investigations are clearly
needed.

Despite recent progress there are many open questions even for the simple
minimal model studies here. For example, the determination of the full phase
diagram is still an open issue both theoretical and numerical. As we have
shown here, in the mean-field picture already the instability line has a
richer structure than anticipated previously. It will be particularly
interesting to study speeds close to $v_\ast$ in order to determine the nature
of the disorder-order transition. However, simulations in this parameter
region are hampered by the large critical fluctuations and will require a more
detailed study of finite-size effects than available so far. Another open
issue is the relation between effective free energy and mechanical
pressure~\cite{taka14,gino15,taka15,bial15,solo14a}.

Finally, while here we have studied analytically the simplest version of
active Brownian particles (namely monodisperse repulsive disks) in two
dimensions without alignment and without hydrodynamic interactions, we note
that multiple extensions of this model have been discussed in the literature:
mixtures of active and passive particles~\cite{sten15}, role of attractive
interactions~\cite{redn13a,mogn13} and alignment~\cite{das14}, and
polydispersity in connection with the glass transition~\cite{ran13,bert14}.


\acknowledgments

We gratefully acknowledge support by DFG within priority program SPP 1726
(grant numbers SP 1382/3-1, ME 3571/2-1, and LO 418/17-1). TS thanks
F. Schmid, P. Virnau, and J. Tailleur for helpful discussions and comments.


\appendix

\section{Run-and-tumble particles}
\label{sec:rnt}

The equivalence of active Brownian particles and run-and-tumble particles on
the level of the hydrodynamic equations has been discussed in
Ref.~\citenum{cate13}. For completeness, we briefly sketch the derivation for
run-and-tumble particles following the route taken in
Sec.~\ref{sec:mf}. Again, we consider $N$ identical particles moving in two
dimensions. A particle moves with constant speed $v_0$ along its orientation
$\vec e\equiv(\cos\vhi,\sin\vhi)^T$ (the ``run''). After an exponentially
distributed random run time with mean $\tu$, the particle ``tumbles'',
\emph{i.e.} it picks a random new orientation. The evolution of the one-point
density $\psi_1(\x,\vhi,t)$ is thus given by
\begin{equation}
  \label{eq:rnt}
  \partial_t\psi_1 = -\nabla\cdot[\mu_0\vec F+v_0\vec e\psi_1] -
  \frac{1}{\tu}\psi_1 + \frac{1}{2\pi\tu}\rho,
\end{equation}
where $\vec F$ is the force due to interactions with other particles, $\mu_0$
the bare mobility, and $\rho(\x,t)$ the local number density
[Eq.~\eqref{eq:rho}]. The second term describes the ``death'' of particles
with a given orientation $\vhi$ and the third term their rebirth with
uniformly distributed orientation. Dimensionless quantities are introduced
through measuring energies in units of $\epsilon$, rescaling time
$t\mapsto\tu t$, and length $\x\mapsto\ell\x$ with
$\ell\equiv\sqrt{\epsilon\mu_0\tu}$.

As a closure, we decompose the force
$\vec F\approx-\rho\zeta\vec e\psi_1-\De\nabla\psi_1$, where the first term
captures the force imbalance along the orientation slowing down the
particles. Since this force is not exactly aligned with the orientation there
will also be an ``evasive'' motion, leading to an effective diffusion on
larger scales. This is described through the second term. Inserting the force
into Eq.~\eqref{eq:rnt}, we thus arrive at [cf. Eq.~\eqref{eq:mf:joint}]
\begin{equation}
  \partial_t\psi_1 = -\nabla\cdot[v(\rho)\vec e-\De\nabla]\psi_1 - \psi_1 +
  \frac{\rho}{2\pi}
\end{equation}
with effective speed $v(\rho)\equiv v_0-\rho\zeta$. It is now straightforward
to derive the hydrodynamic equations~\eqref{eq:rho} and \eqref{eq:p}, which
demonstrates the equivalence of active Brownian particles and run-and-tumble
particles (at least for weakly perturbed orientations).

\section{Differential equation for instability line}
\label{sec:diff}

\begin{figure}[b]
  \centering
  \includegraphics{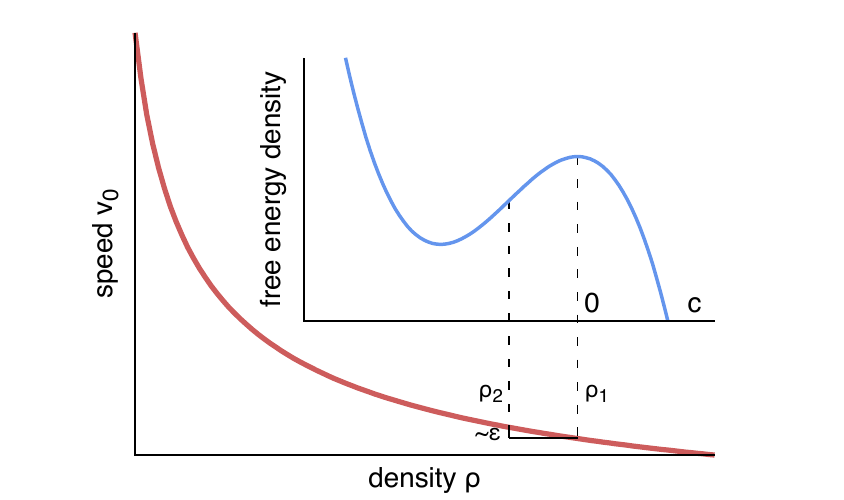}
  \caption{Sketch for the derivation of Eq.~\eqref{eq:cc} for the instability
    line (solid line). For a global density $\rho_1$ the bulk free energy
    density $f(c)$ [Eq.~\eqref{eq:f}] is sketched in the inset, whereby the
    homogeneous density corresponds to $c=0$. Densities down to
    $\rho_2=\rho_1+\eps c\xs_-$ are unstable, which thus corresponds to a
    second point on the instability line.}
  \label{fig:slope}
\end{figure}

The effective free energy density Eq.~\eqref{eq:f} is a local function that
describes the scaled density deviations in the vicinity of a point on the
instability line. Two points along this line have different coefficients
$\sig_1$ and $g$. Still, these cannot be completely independent. Consider a
global density $\rho_1$. Quenching the system to a speed
$v_0=\vc(\rho_1)(1+\eps)$ past the limit of linear stability, separation into
dense and dilute regions will occur. The range of unstable scaled density
deviations is $c\xs_-<c<c\xs_+$, where $c\xs_-=\sig_1/(2g)<0$ is given by the
inflection point of $f(c)$ (the maximal value $c\xs_+$ is irrelevant for the
following argument). Clearly, a suspension with global density
$\rho_2=\rho_1+\eps c\xs_-$ should then become unstable beyond the speed
$v_0=\vc(\rho_2)$. We can exploit this consistency condition to derive a
differential equation that describes the instability line, see
Fig.~\ref{fig:slope} for a sketch of the derivation.

To this end, consider the derivative
\begin{equation}
  \left.\pd{\rho}{v}\right|_{v_1} = \lim_{v_2\ra v_1}
  \frac{\rho_2-\rho_1}{v_2-v_1} = \lim_{v_2\ra v_1}\frac{\eps c\xs_-}{v_2-v_1}
\end{equation}
at speed $v_1=\vc(\rho_1)$. Inserting $\eps=(v_2-v_1)/v_1$, we obtain
\begin{equation}
  \label{eq:cc}
  \pd{\rho}{v} = \frac{\sig_1}{2gv}.
\end{equation}
Knowing one point on the instability line, we can integrate this differential
equation and invert the solution to obtain the full instability line
$\vc(\bar\rho)$, see Ref.~\citenum{spec14} for an example.


\end{document}